\RequirePackage{ifpdf}
\ifpdf 
\documentclass[pdftex]{sigma}
\else
\documentclass{sigma}
\fi

\newcommand{\rmd}{\mathrm{d}}

\begin{document}

\renewcommand{\PaperNumber}{012}

\FirstPageHeading{Yaremko}

\ShortArticleName{Radiation Reaction, Renormalization and
Poincar\'e Symmetry}

\ArticleName{Radiation Reaction, Renormalization \\ and Poincar\'e
Symmetry}

\Author{Yurij YAREMKO} \AuthorNameForHeading{Yu. Yaremko}

\Address{Institute for Condensed Matter Physics of National Academy of Sciences of Ukraine,\\
 1 Svientsitskii  Str., Lviv,  79011 Ukraine}
\Email{\href{mailto:yar@ph.icmp.lviv.ua}{yar@ph.icmp.lviv.ua}}

\ArticleDates{Received July 08, 2005, in final form October 23,
2005; Published online November 01, 2005}

\Abstract{We consider the self-action problem in classical
electrodynamics of a massive point-like charge, as well as of a
massless one. A consistent regularization procedure is proposed,
which exploits the symmetry properties of the theory. The
radiation reaction forces in both 4D and 6D are derived. It is
demonstrated that the {\it Poincar\'e-invariant} six-dimensional
electrodynamics of the massive charge is renormalizable theory.
Unlike the massive case, the rates of radiated energy-momentum
tend to infinity whenever the source is accelerated. The external
electromagnetic fields, which do not change the velocity of the
particle, admit only its presence within the interaction area. The
effective equation of motion is the equation for eigenvalues and
eigenvectors of the electromagnetic tensor. The interference part
of energy-momentum radiated by two massive point charges
arbitrarily moving in flat spacetime is evaluated. It is shown
that the sum of work done by Lorentz forces of charges acting on
one another exhausts the effect of combination of outgoing
electromagnetic waves generated by the charges.}


\Classification{70S10; 78A40}

\section{Introduction}
In classical electrodynamics particles interact with one another
through the medium of a field, which has its own uncountable
infinite degrees of freedom. The dynamics of entire system is
governed by the action
\begin{gather}\label{S}
I=\sum\limits_{a=1}^2\left(I_{a,{\rm part}}+ e_a\int \rmd\tau_a
A_\mu{\dot z}_a^\mu\right) -\frac{1}{16\pi}\int \rmd^4 y
f_{\mu\nu}f^{\mu\nu},
\end{gather}
which is invariant under ten infinitesimal transformations
(space-time translations and rotations) that constitute the
Poincar\'e group. These symmetry properties imply the conservation
laws, i.e.\ those quantities that do not change with time. (In
4-dimensional Minkowski space $I_{a,{\rm part}}$ is proportional
to the worldline length, while in six dimensions the particle part
of action (\ref{S}) involves the term which is proportional to the
curvature of particle's world line \cite{Kos,Nest}.)

\begin{figure}[th]
\centerline{\epsfclipon \epsfig{file=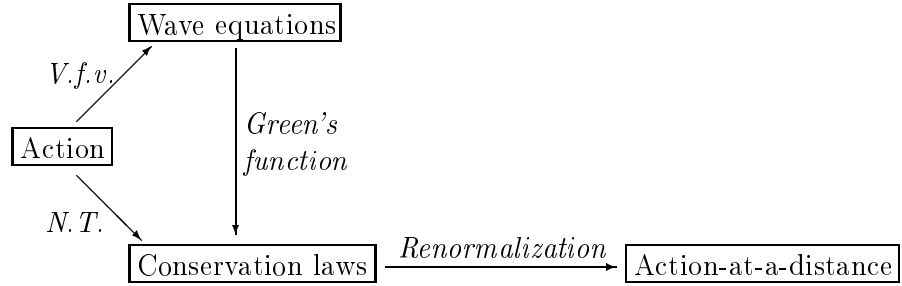,width=12cm}}
\vspace{-1mm}

\caption{\label{gen}We start with standard action principle used
in classical field theory. Variation with respect to field
variables (V.f.v.) yields wave equations. The Poincar\'e
invariance assures us, via Noether's theorem (N.T.), of
conservation laws. The solutions of wave equations with point-like
sources are substituted for field variables in these quantities,
which do not change with time. Having performed renormalization
procedure we derive the effective equations of motion which
involve particles' characteristics only.} \vspace{-3mm}
\end{figure}

Since the field $f_{a,\mu\nu}:=\partial_\mu A_{a,\nu}-\partial_\nu
A_{a,\mu}$ generated by charge $e_a$ has a singularity on its
world line, demanding that the total action \eqref{S} be
stationary under a variation $\delta z_a^\mu(\tau_a)$ of the world
line does not give sensible motion equations. To make sense of the
retarded field's action on the particle, we should perform the
regularization procedure. It can be accomplished via substitution
of Li\'enard--Wiechert solutions for field variables in ten
conserved quantities corresponding to the symmetry of the
action~\eqref{S} under the Poincar\'e group (see Fig.~\ref{gen}).
The conservation of the energy-momentum carried by the
electromagnetic field~\cite{Rohr}
\begin{gather*} 
p_{\rm em}^\nu (\tau)=\int_{\Sigma} \rmd\sigma_\mu T^{\mu\nu}
\end{gather*}
and the angular momentum tensor of the electromagnetic field
\cite{Rohr}
\begin{gather*} 
M_{\rm em}^{\mu\nu}(\tau)=\int_{\Sigma}
\rmd\sigma_\alpha\left(x^\mu T^{\alpha\nu}-x^\nu
T^{\alpha\mu}\right)
\end{gather*}
corresponds to the symmetry of action (\ref{S}) with respect to
space-time translations and space-time rotations, respectively.
(By $\hat T$ we denote the electromagnetic field's stress-energy
tensor; $\rmd\sigma_\mu$ is the outward-directed surface element
on an arbitrary space-like hy\-per\-sur\-fa\-ce $\Sigma$.)

Noether quantity $G^\alpha_{\rm em}$ carried by electromagnetic
field consists of terms of two quite different types \cite{Teit}:
(i) bound, $G^\alpha_{\rm bnd}$, which are permanently
``attached'' to the sources and carried along with them; (ii)
radiative, $G^\alpha_{\rm rad}$, which detach themselves from the
charges and lead independent existence. Within regularization
procedure the bound terms are coupled with energy-momentum and
angular momentum of ``bare'' sources, so that already renormalized
characteristics $G^\alpha_{\rm part}$ of charged particles are
proclaimed to be finite. Noether quantities, which are properly
conserved, become:
\begin{gather*}
G^\alpha=G^\alpha_{\rm part}+G^\alpha_{\rm rad}.
\end{gather*}
In the present paper we apply the regularization scheme based on
the balance equations ${\dot G}^\alpha=0$ to the following
problems:
\begin{itemize}
\itemsep=0pt \item self-action in four- and six-dimensional
classical electrodynamics; \item interference in the radiation of
two point-like sources in 4-dimensional Minkowski space; \item
radiation reaction for a photon-like charge in response to the
external electromagnetic field.
\end{itemize}

\section{On the regularization procedure in classical electrodynamics}

We do not need assumptions concerning with the structure of
singularity of the Maxwell energy-momentum tensor density $\hat T$
on a particle trajectory $\zeta:{\mathbb R}\to {\mathbb M}_{\,4}$.
Our consideration is founded on the energy-momentum and angular
momentum balance equations for a closed system of accelerated
charge $e$ coupled with electromagnetic field. They constitute the
following system of differential equations \cite{YAR,YAR1}:
\begin{gather}\label{em}
{\dot p}_{\rm part}^\mu=-\frac23e^2{\ddot z}^2{\dot z}^\mu\\
{\dot z}\wedge p_{\rm part}=-\frac23e^2{\dot z}\wedge{\ddot z}.
\label{am}
\end{gather}
(Symbol $\wedge$ denotes the wedge product; we use an overdot on
$z$ to indicate differentiation with respect to the evolution
parameter $\tau$.) The solution of equation \eqref{am} involves an
arbitrary scalar function, say $m$:
\begin{gather}\label{Teit}
p_{\rm part}^\mu=m(\tau){\dot z}^\mu-\frac23e^2{\ddot z}^\mu.
\end{gather}
This expression explains how already renormalized particle
4-momentum depends on worldline characteristics, i.e. on
4-velocity ${\dot z}:=u$ and 4-acceleration ${\ddot z}:=a$.

Since ${\dot z}^2=-1$, the scalar product of particle 4-velocity
on the first-order derivative of particle 4-momentum \eqref{em} is
as follows:
\begin{gather}\label{m1}
({\dot p}_{\rm part}\cdot{\dot z})=\frac23e^2{\ddot z}^2.
\end{gather}
Since $({\dot z}\cdot{\ddot z})=0$, the scalar product of particle
acceleration on the particle 4-momentum \eqref{Teit} is given by
\begin{gather}\label{m2}
(p_{\rm part}\cdot{\ddot z})=-\frac23e^2{\ddot z}^2.
\end{gather}
Summing up \eqref{m1} and \eqref{m2}, we obtain
\begin{gather}\label{m}
\frac{\rmd (p_{\rm part}\cdot{\dot z})}{\rmd\tau}=0.
\end{gather}
According to \cite{KosPr}, the scalar product of the 4-momentum
and the 4-velocity $(p_{\rm part}\cdot{\dot z})=m(\tau)$ is the
rest mass of the particle. Since \eqref{m} the function $m$ is of
constant value. Having substituted the expression \eqref{Teit}
with $m={\rm const}$ into equation \eqref{em}, we obtain the
well-known Abraham 4-vector of radiation reaction \cite{Dir}.

\section{Interference in the radiation of two point-like sources}

{\samepage The Maxwell energy-momentum tensor density $\hat T$ is
quadratic in electromagnetic field strengths $\hat f={\hat
f}_{(1)}+{\hat f}_{(2)}$ where ${\hat f}_{(a)}$ denotes the
retarded Li\'enard--Wiechert solution associated with the $a$-th
particle. Hence $\hat T$ can be decomposed as follows \cite{AB}:
\begin{gather*}
T^{\mu\nu} = T_{(1)}^{\mu\nu}+T_{(2)}^{\mu\nu}+T_{\rm
int}^{\mu\nu}.
\end{gather*}
By ${\hat T}_{(a)}$ we mean the contribution due to the field of
the $a$-th particle \cite{Rohr}
\begin{gather*}
4\pi T^{\mu\nu}_{(a)} = f^{\mu\lambda}_{(a)}f^\nu_{(a)\lambda} -
\frac 14
\eta^{\mu\nu}f_{(a)}^{\kappa\lambda}f_{\kappa\lambda}^{(a)}
\end{gather*}
while interference term
\begin{gather}\label{T12}
4\pi T_{\rm int}^{\mu\nu} =
f_{(1)}^{\mu\lambda}f_{(2)}^\nu{}_\lambda +
f_{(2)}^{\mu\lambda}f_{(1)}^\nu{}_\lambda - \frac 14 \eta^{\mu\nu}
\left( f_{(1)}^{\kappa\lambda}f_{\kappa\lambda}^{(2)} +
f_{(2)}^{\kappa\lambda}f_{\kappa\lambda}^{(1)} \right)
\end{gather}
describes the {\em combination} of outgoing electromagnetic
waves.}

Aguirregabiria and Bel \cite{AB} prove the fundamental theorem
that the ``mixed'' radiation rate does not depend on the shape of
the space-like surface, which is used to integrate the Maxwell
energy-momentum tensor density. To integrate the ``mixed''
radiation rate
\begin{gather*} 
p_{\rm int}^\nu = \int_{\Sigma} \rmd\sigma_\mu T_{\rm
int}^{\mu\nu} ,
\end{gather*}
we use the hyperplane $\Sigma_t=\{y\in {\mathbb M}_{\,4}: y^0=t\}$
associated with an unmoving inertial observer. The ``laboratory''
time $t$ is a single common parameter defined along all the world
lines $\zeta_a:{\mathbb R}\to {\mathbb M}_{\,4}$ of the system.
Since the delay in action ensures shifts in arguments in the
electromagnetic field strengths, $a$-th world line $\zeta_a$ is
parametrized by ``individual'' retarded time $t_a$.

The key to the problem is that the angle integration of the
interference contribution (\ref{T12}) results the sum of partial
derivatives in time variables:
\begin{gather*}
\int_0^{2\pi}\rmd\varphi T_{\rm int}^{0\nu}=\frac{\partial^2
G_0^\nu}{\partial t_1\partial t_2}+\frac{\partial
G_1^\nu}{\partial t_1}+ \frac{\partial G_2^\nu}{\partial t_2}.
\end{gather*}
It allows us to calculate the interference part of energy-momentum
carried by two-particle electromagnetic field. It consists of the
bound terms and radiative terms \cite{YAR3}:
\begin{gather*}
p_{\rm int}^\nu =p_{\rm int,bnd}^\nu -\sum_{b\ne
a}\int_{-\infty}^{t}\rmd t_aF_{ba}^\mu
\end{gather*}
The bound terms are absorbed by particles' 4-momenta within the
renormalization procedure.

The radiative part of energy-momentum carried by ``two-particle''
field consists of the integrals of individual Larmor relativistic
rates over corresponding world lines and the work done by Lorentz
forces of point-like charges acting on one another
\cite{YAR2,YAR3}:
\begin{gather}\label{Ptot}
P^\mu=\sum_{a=1}^2\left[p_{a,{\rm part}}^\mu(t)+
\frac23e_a^2\int_{-\infty}^t \rmd t_a(a_a\cdot
a_a)u_a^\mu(t_a)\right] -\sum_{b\ne a}\int_{-\infty}^t \rmd t_a
F_{ba}^\mu.
\end{gather}
Here $p_{a,{\rm part}}$ denotes (already renormalized)
four-momentum of the $a$-th charged particle. The integral of
Larmor relativistic rate describes the contributions ${\hat
T}_{(a)}$  due to the $a$-th individual field, while the sum of
work done by Lorentz force due to the $b$-th particle acting on
the $a$-th one expresses the joint contribution  ${\hat T}_{\rm
int}$ due to combination of fields.

Having differentiated equation (\ref{Ptot}), we arrive at the
relativistic generalization of Newton's second law
\begin{gather*}
{\dot p}_{a,{\rm part}}^\mu=-\frac{2}{3}e_a^2(a_a\cdot a_a)u_a^\mu
+F_{ba}^\mu,
\end{gather*}
where loss of energy due to radiation is taken into account.

\section{Radiation reaction and renormalization in six-dimensional \\
classical electrodynamics}

The Li\'enard--Wiechert potential in six dimensions depends on
particle acceleration \cite{Kos}; in mostly plus signature
\begin{gather*}
A_\mu=e\left[\frac{a_\mu(\tau)}{r^2}+
\frac{u_\mu(\tau)}{r^3}\left(1+ra_k\right)\right] .
\end{gather*}
Here $r$ is the retarded distance \cite{Rohr} and $a_k=a_\alpha
k^\alpha$ is the component of the particle acceleration in the
direction of $k^\alpha := r^{-1}(y^\alpha -z^\alpha(\tau))$.

We assume that an intrinsic structure of singularity of $\hat T$
in the immediate vicinity of particle's world line $\zeta:{\mathbb
R}\to {\mathbb M}_{\,6}$ is beyond the limits of classical theory.
The radiative part of electromagnetic field's energy-momentum is
as follows \cite{YAR6D,YAR6}:
\begin{gather} \label{prad}
p_{\rm rad}^\mu=e^2 \int_{-\infty}^t \rmd\tau
\left(\frac45(\dot{a}\cdot\dot{a})u^\mu-\frac{6}{35}(a\cdot
a)\dot{a}^\mu +\frac37a^\mu (a\cdot a)^{\bf\cdot}+2(a\cdot
a)^2u^\mu\right),
\end{gather}
where $(a\cdot a)^{\bf\cdot}$ denotes the derivative of the square
of particle acceleration with respect to evolution parameter
$\tau$.

The radiative part of electromagnetic field's angular momentum
depends on all previous motion of a source up to the instant of
observation $t$ \cite{YAR6D,YAR6}:
\begin{gather}
M_{\rm rad}^{\mu\nu}=e^2 \left\{ \int_{-\infty}^{t} \rmd\tau
\left( z^\mu P_{\rm rad}^\nu - z^\nu P_{\rm rad}^\mu \right)
\right.
\nonumber\\
\left.\phantom{M_{\rm rad}^{\mu\nu}=}{}+ \int_{-\infty}^{t}
\rmd\tau \left[ \frac45\left(a^\mu{\dot a}^\nu-a^\nu{\dot
a}^\mu\right) +\frac{64}{35}(a\cdot a)\left(u^\mu a^\nu-u^\nu
a^\mu\right)\right]\right\}. \label{Mrad}
\end{gather}
Here $P_{\rm rad}$ denotes the integrand of equation~\eqref{prad}.

With \eqref{Mrad} in mind we assume that already renormalized
angular momentum tensor of the particle has the form
\begin{gather*} 
M_{\rm part}^{\mu\nu} =z^\mu p_{\rm part}^\nu-z^\nu p_{\rm
part}^\mu + u^\mu\pi_{\rm part}^\nu -u^\nu\pi_{\rm part}^\mu .
\end{gather*}
In \cite{Nest} the extra momentum $\pi_{\rm part}$ is due to
additional degrees of freedom associated with acce-le\-ra\-tion
involved in Lagrangian function for rigid particle.

Scrupulous analysis of consistency of the energy-momentum balance
equation and angular momentum balance equation reveals that
six-momentum of charged particle contains two (already
renormalized) constants, $m$ and $\mu$:
\begin{gather} \label{p_ost}
p_{\rm part}^\beta = mu^\beta +\mu\left(-{\dot a}^\beta
+\frac32(a\cdot a)u^\beta\right) +e^2\left[\frac45{\ddot a}^\beta
-\frac85u^\beta (a\cdot a)^{\bf\cdot}-\frac{64}{35}(a\cdot
a)a^\beta\right].
\end{gather}

Having substituted the right-hand side of \eqref{p_ost} for the
particle's six-mo\-men\-tum in the energy-mo\-men\-tum balance
equation ${\dot p}_{\rm part}+{\dot p}_{\rm rad}=0$, we derive the
Lorentz--Dirac equation of motion of a charged particle under the
influence of its own electromagnetic field. An external device
adds covariant external force $F_{\rm ext}$ to the right-hand side
of this expression \cite{Kos}:
\begin{gather*} 
{\dot p}_{\rm part}^\mu +e^2
\left(\frac45u^\mu\dot{a}^2-\frac{6}{35}a^2\dot{a}^\mu
+\frac37a^\mu (a^2)^\cdot+2a^4u^\mu\right)=F_{\rm ext}^\mu .
\end{gather*}

\section{Radiation reaction and renormalization for a photon-like \\
charged particle}

Our consideration is based on the Maxwell equations with
point-like source \cite[equation~(14)]{KS}
\begin{gather*}
\Box A_\mu (x)=-4\pi j_\mu (x),
\end{gather*}
which governs the propagation of the electromagnetic field
produced by a photon-like charge. Current density is zero
everywhere, except at the particle's position where it is infinite
\begin{gather*}
j_\mu (x)=q\int\rmd\tau u_\mu (\tau)\delta[x-z(\tau)]
\end{gather*}
and $\Box :=\eta^{\alpha\beta}\partial_\alpha\partial_\beta$ is
the wave operator. Unlike the massive case, the photon-like
charge, say $q$, generates the {\it far} electromagnetic field
\begin{gather}\label{f}
\hat f=q\frac{a\wedge k+a_ku\wedge k}{r},
\end{gather}
which does not yield to divergent Coulomb-like self-energy. Hence
the photon-like charge does not possess an electromagnetic
``cloud'' permanently attached to it. The renormalization
procedure is not necessary because the photon-like source is not
``dressed''.

As a consequence, the Brink--Di~Vecchia--Howe action term
\cite{BVH}:
\begin{gather}\label{prt}
I_{\rm part} = \frac12\int \rmd\tau e(\tau){\dot z}^2
\end{gather}
is consistent with the field and the interaction terms (\ref{S}).
Variation of (\ref{prt}) with respect to Lagrange multiplier
$e(\tau)\ne 0$ yields the isotropy condition ${\dot z}^2=0$. The
particle part (\ref{prt}) of the total action (\ref{S}) describes
{\it already renormalized} massless charge.

The retarded distance involved in equation~(\ref{f})
\begin{gather}\label{r}
r=-\eta_{\alpha\beta}[x^\alpha -z^\alpha(s)]u^\beta(s)
\end{gather}
contains particle's characteristics (position, $z^\alpha$, and
4-velocity, $u^\beta$) taken at the retarded time,~$s$, being the
(causal) root of algebraic equation
\begin{gather*}
\eta_{\alpha\beta}[x^\alpha -z^\alpha(s)][x^\beta -z^\beta(s)]=0.
\end{gather*}
Because of the isotropy condition, the retarded distance (\ref{r})
vanishes on the ray in the direction of particle's 4-velocity
taken at the instant of emission $z(s)\in\gamma$. Unlike the massive case, the
field (\ref{f}) diverges at all the points of this ray with vertex
at the point of emission.

Volume integration of the radiative energy-momentum tensor density
\begin{gather*}
4\pi T^{\alpha\beta}=\frac{q^2}{r^2}a^2k^\alpha k^\beta
\end{gather*}
over a hyperplane $\Sigma_t=\{x\in {\mathbb M}_{\,4}: x^0=t\}$
gives the amount of radiated energy-momentum at fixed instant $t$.
We introduce the set of curvilinear coordinates for flat
space-time ${\mathbb M}_{\,4}$ involving the observation time $t$
and the retarded time $s$:
\begin{gather*}
x^\alpha=z^\alpha(s) + (t-s)\Omega^\alpha{}_{\alpha'}n^{\alpha'}.
\end{gather*}
The null vector $n:=(1,{\boldsymbol n})$ has the components
$(1,\cos\varphi\sin\vartheta,\sin\varphi\sin\vartheta,\cos\vartheta)$;
$\vartheta$ and $\varphi$ are two polar angles which distinguish
the points on the spherical wave front
\begin{gather*} 
S(z(s),t-s)=\left\{x\in {\mathbb M}_{\,4}:
(x^0-s)^2=\sum_i(x^i-z^i(s))^2,x^0=t\right\},
\end{gather*}
which is the intersection of the future light cones with vertex at
point $z(s)\in\gamma$ and hyperplane~$\Sigma_t$. Matrix space-time
components are $\Omega_{0\mu}=\Omega_{\mu 0}=\delta_{\mu 0}$; its
space components $\Omega_{ij}$ constitute the orthogonal matrix
which rotates space axes of the laboratory Lorentz frame until new
$z$-axis is directed along three-vector ${\boldsymbol v}$.
(Particle's 4-velocity has the form $(1,v^i)$, $|{\boldsymbol
v}|=1,$ if parameterization of the world line $\gamma$ is provided
by a disjoint union of hyperplanes $\Sigma_t$.) In terms of
curvilinear coordinates $(t,s,\vartheta,\varphi)$ the retarded
distance (\ref{r}) is as follows:
\begin{gather*}
r=(t-s)(1-\cos\vartheta).
\end{gather*}

\begin{figure}[th]
\centerline{\epsfclipon \epsfig{file=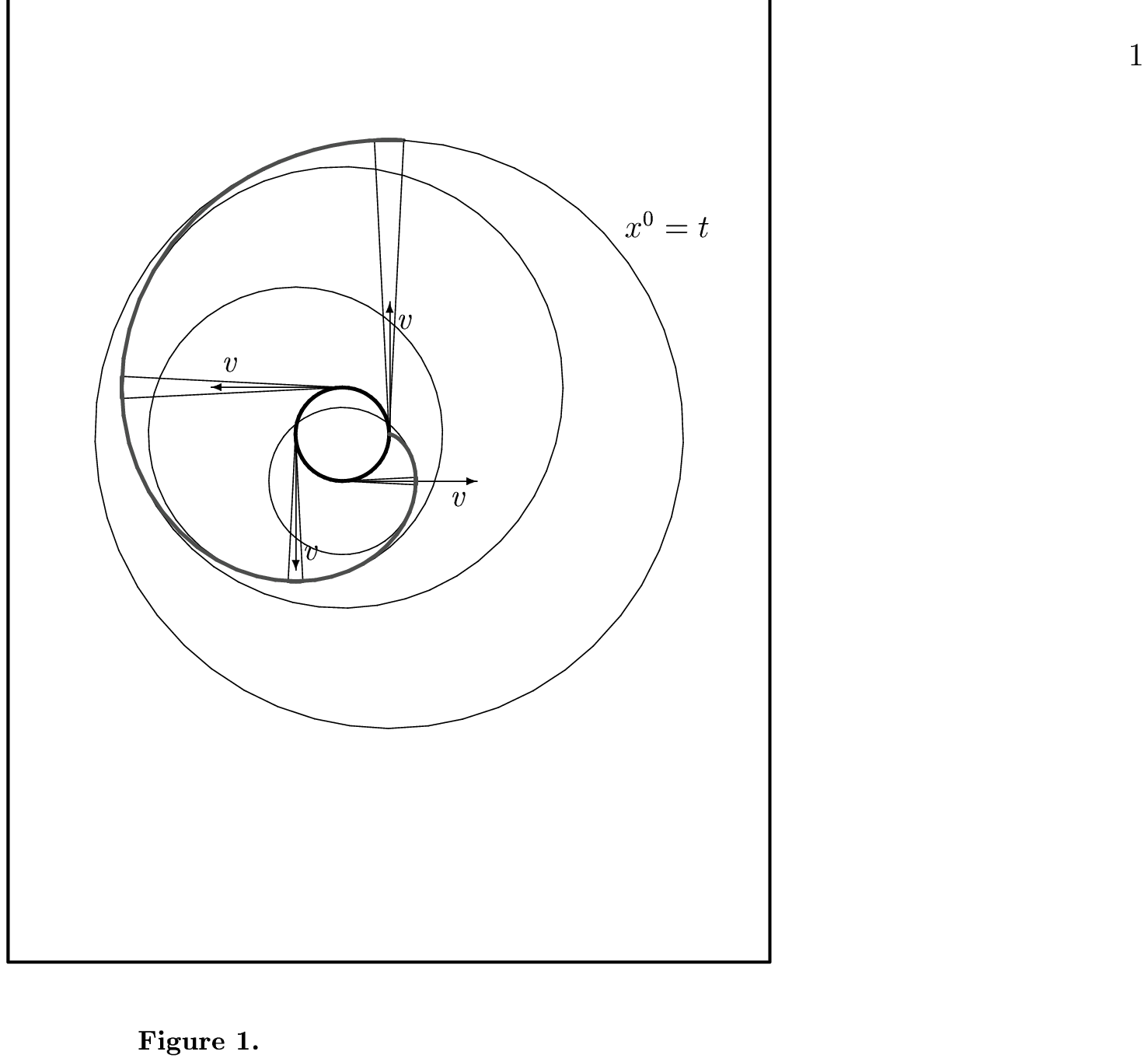,width=8cm}}
\vspace{-1mm} \caption{\label{rad} The bold circle pictures the
trajectory of a photon-like charge. The others are spherical wave
fronts viewed in the observation hyperplane $\Sigma_t=\{x\in
{\mathbb M}_{\,4}: x^0=t\}$. The circling photon-like charge
radiates infinite rates of energy-momentum and angular momentum in
the direction of its velocity ${\boldsymbol v}$ at the instant of
emission. The energy-momentum carried by electromagnetic field of
accelerated charge tends to infinity on the spiral curve.}
\vspace{-3mm}
\end{figure}

Angle integration reveals that the Larmor term tends to infinity
whenever the photon-like charge is accelerated:
\begin{gather*}
p^0_{\rm em}=\frac{q^2}{2}\left(
-\frac{1}{8}+\lim\limits_{\vartheta\to 0}
\frac{1}{2(1-\cos\vartheta)^2}\right) \int_{-\infty}^t\rmd\tau
a^2(\tau),
\\
p^i_{\rm em}=\frac{q^2}{2}\left(
\frac{3}{8}-\lim\limits_{\vartheta\to 0}
\left[\frac{1}{1-\cos\vartheta}-\frac{1}{2(1-\cos\vartheta)^2}\right]\right)
\int_{-\infty}^t\rmd\tau a^2(\tau)v^i(\tau).
\end{gather*}
The situation is pictured in Fig.~\ref{rad}. Since the emitted
radiation detaches itself from the charge and leads an independent
existence, it cannot be absorbed within a renormalization
procedure.

To change the velocity of the massless charge the energy is
necessary which is too large to be observed, while non-accelerated
photon-like charge does not generate the electromagnetic field.
The evolution of the particle beyond an interaction area is
determined by the Brink--Di~Vecchia--Howe Lagrangian~(\ref{prt}).
 The particle's 4-momentum $p_{\rm part}^\mu=e(\tau){\dot z}^\mu$ does
not change with time: ${\dot e}(\tau){\dot z}^\mu=0$. Since ${\dot
z}^\mu\ne 0$, the Lagrange multiplier $e$ does not depend on
$\tau$. We deal with a photon-like particle moving in the $
\dot{\boldsymbol z}$-direction with frequency $\omega_0=e_0{\dot
z}^0$, such that its energy-momentum 4-vector can be written
$p_{\rm part}^\mu=(\omega_0,\omega_0{\boldsymbol k})$,
$|{\boldsymbol k}|=1$.

When considering the system under the influence of an external
device, the change in particle's 4-momentum is balanced by the
external force:
\begin{gather} \label{ef}
{\dot e}v^\mu=qF^\mu{}_\nu v^\nu.
\end{gather}
This effective equation of motion is supplemented with the
condition of absence of radiative damping. In other words, the
external device admits a massless charge if and only if the
components of null vector of 4-velocity do not change with time
despite the influence of the external field.

The expression (\ref{ef}) is obtained in \cite{Ryl} where the
model of magnetosphere of a rapidly rotating neutron star (pulsar)
is elaborated. (The gas of ultrarelativistic electrons and
positrons moving in a very strong electromagnetic field of the
pulsar is meant.) If the gradient of star's potential is much
larger than the particle's rest energy $m_ec^2$, the strong
radiation damping suppresses the particle gyration. It is
shown~\cite{Ryl} that the particle velocity is directed along one
of the eigenvectors of (external) electromagnetic tensor $\hat F$
if $m_e\to 0$ in the Lorentz--Dirac equation. Equation (\ref{ef})
on eigenvalues and eigenvectors of the electromagnetic tensor
governs the motion of charges in zero-mass approximation. This
conclusion is in contradiction with that of Ref.~\cite{KS}, where
the radiation back reaction is finite and the 5-th order
differential equation determines the evolution of photon-like
charge. The reason is that the Dirac regularization approach to
the radiation back reaction (one-half retarded field minus
one-half advanced one), employed by Kazinski and Sharapov, is not
valid in the case of the photon-like charged particle and its
field. Indeed, the advanced field is free from the ray
singularity, and the Dirac combination does not yield a finite
part of the self force.

\subsection{Conformal invariance}
Action integral (\ref{S}) with $I_{\rm part}$ in form of
(\ref{prt}) is conformally invariant. According to \cite{FN},
conformal group ${\cal C}(1,3)$ consists of Poincar\'e
transformations (time and space translations, space and mixed
space-time rotations), dilatations
\begin{gather}\label{dil}
x'{}^{\mu}=e^\theta x^\mu
\end{gather}
and conformal transformations
\begin{gather}\label{ct}
x'{}^{\mu}=\frac{x^\mu-b^\mu (x\cdot x)}{D},\qquad D=1-2(x\cdot b)
+ (x\cdot x)(b\cdot b).
\end{gather}
(The scalar $\theta$ and 4-vector $b$ are group parameters.)

The components of electromagnetic field are transformed as
follows:
\begin{gather*}
F_{\alpha\beta}=e^{2\theta}F'_{\alpha\beta}, \qquad
F_{\alpha\beta} = F'_{\mu\nu}\Omega^\mu{}_\alpha
\Omega^\nu{}_\beta
\end{gather*}
where matrix
\begin{gather*}
\Omega^\mu{}_\alpha:=\frac{\partial x'{}^{\mu}}{\partial
x^\alpha}=
D^{-1}\lambda^\mu{}_\beta(x'')\lambda^\beta{}_\alpha(x),\qquad
x''=\frac{x}{(x\cdot x)}-b,\qquad
\lambda^\beta{}_\alpha(x)=\delta^\beta{}_\alpha-\frac{2x^\beta
x_\alpha}{(x\cdot x)}
\end{gather*}
satisfies the condition
\begin{gather*}
\eta_{\mu\nu}\Omega^\mu{}_\alpha\Omega^\nu{}_\beta=D^{-2}\eta_{\alpha\beta}
.
\end{gather*}
Since
\begin{gather*}
{\dot z}^{'\mu} =e^\theta {\dot z}^\mu \qquad {\dot z}^{'\mu} =
\Omega^\mu{}_\alpha {\dot z}^\alpha
\end{gather*}
the Lagrange multiplier $e(\tau)$ involved in the
Brink--Di~Vecchia--Howe action term (\ref{prt}) transforms as
\begin{gather*}
e(\tau)=e^{2\theta}e'(\tau),\qquad e(\tau)=D^{-2}e'(\tau).
\end{gather*}
Direct calculation shows that the effective equation of motion
(\ref{ef}) is invariant with respect to dilatation (\ref{dil}) and
conformal transformation (\ref{ct}).

It is worth noting that the conformal invariance yields
conservation laws, which are functions of energy-momentum and
angular momentum conserved quantities.

Conformal invariance of our particle plus field system reinforces
our conviction that the back-reaction force vanishes. Indeed, the
appropriate renormalization procedure should preserve this
symmetry property while the the Brink--Di~Vecchia--Howe action
term (\ref{prt}) does not contain a~parameter to be renormalized.
Therefore, the photon-like charge must not radiate.

\section{Conclusions}
The law of conservation of the total four-momentum of a composite
(particle plus field) system provides the foundation for Dirac's
derivation \cite{Dir} of the radiation-reaction force. It involves
the Taylor expansion of a finite sized charged sphere, in which
the first two terms lead to the electromagnetic self-energy, and
the Abraham radiation reaction four-vector, respectively.
Inevitable infinities arising in six-dimensional electrodynamics
are stronger than in four dimensions. In~\cite{Kos} Kosyakov
calculates the flux of energy-momentum and derives the
radiation-reaction force in six-dimensions by adding of
appropriate Schott term.

In the present paper we consider also the conserved quantities
corresponding to the invariance of the theory under proper
homogeneous Lorentz transformations. They gives additional
information which allows us to avoid the problems concerning with
an intrinsic structure of a~singularity. In such a way we
reformulate the problem of renormalizability within the problem of
Poincar\'e invariance of a closed particle plus field system. The
conservation laws are an immovable fulcrum about which tips the
balance of truth regarding renormalization and radiation reaction.

The renormalization scheme is applied to the problem of
combination of outgoing electromagnetic waves generated by two
point-like sources \cite{YAR2,YAR3}. The evaluation of the
interference part of energy-momentum radiated by charges is not a
trivial matter, since delay in action ensures shifts in arguments
in the electromagnetic field strengths. Rigorous calculations show
that the interference rate of radiation which escapes to infinity
is equal to the rate of work done by Lorentz forces of charges
acting on one another. The bound terms arise, which can be
interpreted as a usual deformation of the bound electromagnetic
``clouds'' of charges due to mutual interaction. They are absorbed
within the renormalization procedure, as well as the inevitable
infinities arising in a one-particle problem.

Inspection of the energy-momentum and angular momentum carried by
the ele\-ctro\-mag\-ne\-tic field of a photon-like charge reveals
the reason why it is not yet detected (if it exists).
Noninteracting massless charges do manifest themselves in no way.
Any external electromagnetic field (including that generated by a
detecting device) will attempt to change the velocity of the
charged particle. Whenever the effort will be successful, the
radiation reaction will increase extremely. In general, this
circumstance forbids the presence of the photon-like charges
within the interaction area.

To survive, photon-like charges need an extremely strong field of
specific con\-fi\-gu\-ra\-ti\-on, as that of the rotating neutron
star (pulsar). In \cite{Ryl} the model of the pulsar magnetosphere
is elaborated. It involves the so-called {\it dynamical phase},
which consists of the massless charged particles moving with speed
of light along real eigenvalues of the electromagnetic field
tensor of the star.

\subsection*{Acknowledgements}
The author would like to thank Professors B.P.~Kosyakov,
V.~Tre\-tyak, Dr.~A.~Du\-vi\-ryak, and Professor Yu.A.~Rylov for
helpful discussions and critical comments.

\LastPageEnding

\end{document}